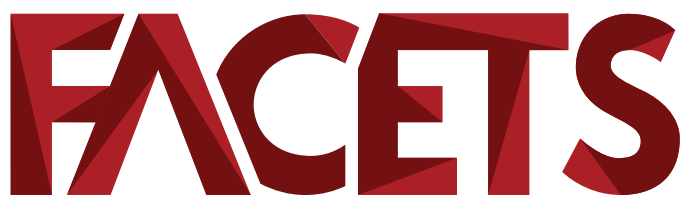

# Canadian physics counts: an exploration of the diverse identities of physics students and professionals in Canada

**Eden J. Hennessey**[a], **Anastasia Smolina**[b], **Skye Hennessey**[a], **Adrianna Tassone**[a], **Alex Jay**[a], **Shohini Ghose**[c,d], and **Kevin Hewitt**[e]

[a]Wilfrid Laurier University, Department of Psychology, Waterloo, ON N2L 3C5, Canada; [b]University of Toronto, Department of Medical Biophysics, Toronto, ON M5G 2C4, Canada; [c]Wilfrid Laurier University, Department of Physics and Computer Science, Waterloo, ON N2L 3C5, Canada; [d]Quantum Algorithms Institute, Surrey, BC V3T 5X3, Canada; [e]Dalhousie University, Department of Physics and Atmospheric Science, Halifax, NS B3H 4R2, Canada

Corresponding authors: **Eden J. Hennessey** (email: ehennessey@wlu.ca); **Anastasia Smolina** (email: anastasia.smolina@mail.utoronto.ca)

## Abstract

The lack of diversity in physics remains a worldwide problem. Despite being a quantitative discipline relying on measurements, there is a paucity of data on the demographic characteristics of underrepresented groups. We present findings from Canadian Physics Counts: the first national survey of equity, diversity, and inclusion in the Canadian physics community. Our intersectional approach provides a wealth of information on gender identity, sexual orientation, race, disability, and more. Analyses revealed key findings, including the first comprehensive Canadian data on nonbinary or gender diverse physicists, and the first data on Black (1.2%) and Indigenous (0.3%) physicists, who were least represented. Assessing representation across roles and career stages, BIPOC (Black, Indigenous, and People of Colour), representation fell precipitously (by one-half) between under/graduate to professional roles, while White women's representation remained relatively stable, and White men's representation steadily increased. One in four BIPOC respondents from gender diverse backgrounds identified as disabled, and the proportion of sexually diverse students with disabilities was more than three times higher than the proportion of heterosexual students with disabilities, underscoring the necessity of intersectional analyses. Students were more demographically diverse than professionals, highlighting the importance of acting today to retain the diverse physicists of tomorrow

## Introduction

Participation of diverse groups in physics in Canada has been persistently low (Science Secretariat 1967; Robertson and Steintz 1997; Perreault et al. 2018). Demographic information is crucial to understand gaps in representation, and to inform development, implementation, and evaluation of polices and interventions to improve access and participation in physics. Demographic data for physicists has been collected in the United States (e.g., Porter and Ivie 2019; American Institute of Physics 2020), the United Kingdom (The Royal Society 2025), and Australia (Maasoumi et al. 2023), however, data collected in Canada is limited and only exists for binary gender (Science Secretariat 1967; Robertson and Steintz 1997; CAUT 2014; Strickland 2017), making it difficult to understand how diverse identities intersect in the North American context. The purpose of this work was to collect data from the Canadian physics community, applying an intersectional analysis to better understand who is and who is not represented.

Previous research from social and organizational psychology has demonstrated the relevance of race, gender, and 2SLGBTQIA+ identity in determining who studies, works, and succeeds in science. Such factors range from aspects of individual experience, such as identification with science (Robinson et al. 2019), confidence (Sterling et al. 2020), access to resources (Ivie et al. 2014; Porter and Ivie 2019), and racism (American Institute of Physics 2020; Clark and Hurd 2020), to societal factors such as socialization (Tenenbaum and Leaper 2003), stereotypes about scientists (Reuben et al. 2014; Banchefsky and Park 2018), the predominance of Whiteness in the academy (Johnson and Howsam 2020), and the culture and climate of classrooms and labs (Settles et al. 2006; Walton et al. 2015). Research shows that racialized or BIPOC (Black, Indigenous, and People of Colour) experience bias and discrimination such as being underestimated and unrecognized as physicists (Hyater-Adams et al. 2019). Members of the 2SLGBTQIA + community in physics face exclusionary behaviour, discomfort in the workplace, mockery, and harassment (Barthelemy 2020; Barthelemy et al. 2022; Gutzwa et al. 2024), and this may be exacerbated for transgender physicists (Barthelemy et al. 2022).

Social justice (i.e., efforts to create a society wherein all people can access rights, opportunities, and resources, inclusive of demographic identities or circumstances) must be better integrated into science, technology, engineering and mathematics (STEM) to attract those who have faced social injustice. Gibbs and Griffin (2013) found that among post-doctoral scholars, non-URM (underrepresented minority) scientists often cited freedom to choose their own research topics as a main reason to pursue a faculty career, whereas URM post-doctoral scholars were more often driven by goals such as mentoring students like themselves and addressing problems in their communities. Garibay (2015), in a study comprising 6100 undergraduates found that equity, social justice, helping others through their work and working for social change, were more important to URM STEM students than their non-URM peers. McGee (2021) concluded from these studies that "justice-oriented STEM" (i.e., STEM that acknowledges historical and contemporary inequities), can be a critical factor in attracting and retaining diverse people in science.

To more fully understand experiences and representation in science, some research has drawn on intersectional theory (i.e., recognition that marginalized identities such as gender, race, ability, interact to create greater oppression; Crenshaw 2017). For example, the "double bind", articulates how racism and sexism combine to create significant barriers to the participation of women of colour in science (Malcom et al. 1976; Malcom and Malcom 2011; Ong et al. 2011). Johnson and colleagues (2017) conducted interviews with women of colour in physics who described experiences of isolation, stereotyping, and microaggressions. Similarly, Rosa and Mensah (2016) found that Black women studying physics experienced isolation and exclusion from important resources (e.g., study groups), encountering barriers to participation.

Given the barriers to participation in physics experienced by different demographic groups, it is perhaps unsurprising that women and BIPOC communities are underrepresented. And, among those with multiple, intersecting marginalized identities, underrepresentation is even more pronounced. Ivie and colleagues (2014) found that out of approximately 9000 physics and astronomy faculty members in the United States, there were fewer than 75 women who were either African American or Hispanic. Indeed, BIPOC women are barely represented in physics; in all of history, fewer than 10 Black women have graduated with PhDs in theoretical high energy physics (Prescod-Weinstein 2015). BIPOC women are underrepresented at the PhD level in physics, and racial disparities are well-documented in earlier post-secondary education as well. A recent report from the United States showed a disturbing decline in the participation of African Americans in physics education over decades; in 1990 about 5% of bachelor's degrees in physics were awarded to African Americans and less than 4% in 2017 (American Institute of Physics 2020). Given the lack of representation of BIPOC physicists, and particularly BIPOC women at all academic levels, and the "double bind" challenge of experiencing both persistent racism and sexism, it is imperative to apply an intersectional analysis to research. Yet, there is a lack of disaggregated data on Canadian physicists that has applied an intersectional analysis.

Although Canadian demographics are similar to the United States in some ways, they are very different in others (e.g., African American vs. African Canadian population). However, few studies have focused specifically on the demographic composition of the Canadian physics community. Considering researchers have outlined barriers to achieving equity in Canada's higher education system, including unconscious and implicit bias, affinity bias, White normativity, tokenism, the equity tax, and other factors impacting career progression and lived experiences in academia (Henry et al. 2017; James and Turner 2017), a national survey of the Canadian physics community to analyze the unique intersecting identities of its members is long overdue.

Existing data on physics students and professionals in Canada are sparse, disconnected, and often focus on limited identities (e.g., only gender), making intersectional analyses unlikely or impossible. Historically, a 1967 report by the Canadian Association of Physicists (CAP) headed by D.C. Rose assessed the physics community through a survey that gauged numbers of faculty and students by department. While foundational, this work did not include an analysis of demographic identities, such as gender or race (Science Secretariat 1967). Three decades later, another analysis was conducted of highly qualified personnel who had obtained physics degrees at 52 Canadian universities and colleges (Robertson and Steinitz 1997). However, this report included only one question on respondent sex (i.e., response options were "male" and "female"). National statistics often reflect aggregate data, such as counts of faculty members in physical sciences, computer science, engineering, and math (Council of Canadian Academies 2012), bachelor's students in chemistry and physics (Statistics Canada 2014), but all measure binary sex (i.e., male and female; men and women).

Similarly, an independent survey by the Canadian Association of University Teachers (CAUT) in 2010 reported frequencies of physics faculty members (CAUT 2014). In 2017, Nobel Prize winner Donna Strickland reported on physics departments in Canada, noting proportions of female students and faculty members; however, this analysis did not include any other aspects of demographic identity, such as race (Strickland 2017). A 2018 report showed enrolment of women and visible minorities in undergraduate physics programs from 2000 to 2009 but did not disaggregate data beyond binary sex and race (Perreault et al. 2018). The Canadian Organization of Medical Physicists (COMP) conducted a survey on climate and collected data on race, gender, disability, and other identity variables from respondents. However, this study did not use an intersectional analysis and was limited to medical physicists who were COMP members. Finally, the CAP collected data on gender, race, and disability. However, this data was limited to physics students, faculty, and staff at select universities, and did not include an intersectional analysis. Additionally, the data were based on institutional statistics, where the vast majority did not have data on gender identities beyond men and women (i.e., CAP 2023). In sum, our understanding of the demographics of the physics community in Canada has been largely informed by few anal-yses, selected reports from Statistics Canada (Hango 2013; Wall 2019; Chan et al. 2021), and research focusing on particular identities in physics (e.g., women; Predoi-Cross et al. 2019; Dengate et al. 2021).

**Table 1.** Estimated totals of Canadian physics subgroups.

| Group | Estimated total possible responses | Obtained sample size |
|---|---|---|
| **Undergraduate students** | 3000 | 895 |
| **M.Sc. students** | 500 | 246 |
| **Ph.D. students** | 500 | 388 |
| **Postdoctoral Fellows** | 450 | 115 |
| **Faculty** | 1000 | 397 |
| **Industry + Government** | 11000 | 225 |

In Canada, women are not adequately represented in physical and chemical sciences and in post-secondary physics education, despite comprising the largest proportion of post-secondary graduates (Wall 2019). Some evidence suggests that the transition to college/university represents the largest loss of women from STEM programs—women make up only 34% of Grade 12 physics students in Ontario, which is a required course for post-secondary physics education (Wells et al. 2018). In response to the lack of gender diversity within Canadian science, including in physics, various initiatives to promote participation have been implemented, such as conferences promoting women in physics (Predoi-Cross et al. 2019), the NSERC Chairs for Women in Science and Engineering Program (2021), NSERC Chairs for Inclusion in Science and Engineering (2023) and the Waterloo Charter for Gender Inclusion and Diversity in Physics (International Union for Pure and Applied Physics 2021). Although it is a step in the right direction to understand women's representation and retention in Canadian physics, gender is only one facet of identity that relates to the retention of physics students, who are the future of physics (Porter and Ivie 2019). Despite the importance of assessing the multifaceted identities of scientists, there has never been a national survey to collect detailed demographic data (i.e., gender, race, sexual orientation, and disability) among physics students or faculty in Canada. Without such data to guide decision-making, the physics community cannot meaningfully address gaps in representation or experience.

To this end, in the present study we conducted the first ever Canada-wide survey to assess representation in physics. We focus on two questions: (1) what are the demographic identities of post-secondary physics students and working professionals in Canada? (2) How do the demographic identities of physics post-secondary students (and working professionals) in Canada intersect? Detailed demographic data were gathered from more than 2500 physicists gauging gender identity, sexual orientation, race, disability, and more. This allowed us to explore nuanced details about demographics along multiple axes of identity—to our knowledge a global first. Importantly, we note that demographic data alone is insufficient to understand how to improve participation in physics. However, in concert with other research (i.e., experimental research, case studies) demographic information can be used to inform recruitment strategies and refine policies and interventions aimed at enhancing access and participation in physics.

## Method

### Survey design

This research was approved by the Dalhousie University Research Ethics Board, 20205261. To participate, respondents were required to either (1) be pursuing a physics degree at a Canadian institution, or (2) hold a physics degree and be working/residing in Canada, as of November 2020. This approach was chosen to include a wide array of respondents and capture the experiences of those working beyond academia. Our approach was to divide all respondents into two overarching groups with multiple sub-categories: students (i.e., undergraduate students, graduate students), and working professionals (i.e., postdoctoral fellows, research staff, faculty, or working professionals in government or the private sector). To estimate the number of physicists in each sub-category, the CAP contacted the representative heads of all Canadian institutions offering a physics degree ($N = 60$) requesting data on number of members. Responses were received from 64% of institutions, and used alongside enrolment rates, degree length, and historical data to project estimated total possible responses for all subgroups shown in Table 1. Regarding industry and government positions, given approximately 1100 physics undergraduates graduate annually (Strickland 2017), and approximately 40% begin a career in industry or government (Robertson and Steinitz 1997), assuming an average career length of 25 years yields a total possible cohort of 11 000 physicists employed in the private sector or government. Given these data and assumptions, we projected estimated totals (weighted by degree enrolment and average degree completion time) that were used as a guide to ensure that our obtained sample included physicists from various subgroups. As illustrated in Table 1, students comprised the largest proportion of respondents.

The survey was constructed on a private organizational account owned by the CAP. Privacy and confidentiality were ensured by separating personally identifiable respondent information, and an application programming interface linking system was used to connect individual responses for data analysis. The survey was piloted with a small sample ($n = 24$) for length and averaged 10.4 min to complete the 38–51 ques-tions (range due to branching structure), which fell within the expected range of 10–15 min.

## Survey procedure

Invitations to complete the survey were sent via email to all members of the CAP, every post-secondary physics department in the country, institutional partners (e.g., Perimeter Institute, SNOLAB, TRIUMF, Canadian Light Source, Institute for Quantum Computing, Stewart Blusson Quantum Matter Institute, CAP's network of high school physics teachers, industrial partners, and professional associations (e.g., Canadian Organization of Medical Physicists, Canadian Astronomical Society). All survey respondents were encouraged to share the survey link with any other physics students or physicists in their networks. Therefore, some respondents were recruited utilizing a "snowball method". The survey was available in French and English. French entries (9.8%) were translated into English prior to analysis.

The recruitment email (see Supplementary Materials) described the survey process and emphasized participant confidentiality and anonymity. After providing informed consent, respondents completed various measures assessing demographic characteristics. Respondents were thanked for their time and could enter an email address for a draw (stored separately from survey responses). As participation was voluntary and some respondents were recruited through a snowball method, findings cannot be interpreted as representative of the full physics community in Canada, but rather as a subset of the community who volunteered, who may differ from the population in unknown ways. For example, some identity groups may be more willing to participate in surveys in general (Porter and Umbach 2006; Lee et al. 2018). However, other groups may be less likely to report identity (Dembosky et al. 2019). Self-selection is an unavoidable component of social research. When collecting demographic information, it is unethical and impractical to assume people's identities, making self-report, and therefore consent and self-selection a necessary component of the research. Despite these limitations, this approach was chosen over a CAP member census to mitigate the perception of coercion and to reach a larger portion of the physics community, including nonmembers of the association (which include the majority of undergraduate physics students).

## Materials

The survey asked respondents to provide information regarding employment, education, demographic characteristics, and experiences in physics. Given the scarcity of data on representation of physics students and physicists in Canada, the analysis presented here focuses on demographic variables. In our analysis, some demographic information (e.g., gender identity, race, and sexual orientation), was collapsed into overarching variables; for example, a BIPOC (i.e., Black, Indigenous, People of Colour) category. This practice has been used in past research where race is a variable of interest because it helps maintain anonymity and appropriate statistical power (e.g., Magoon et al. 2022; Centre for Research and Innovation Support 2024; Coleman and Yantis 2024; Jin et al. 2024; Levandowski et al. 2024). Despite its prevalence, we acknowledge that this practice is not ideal, as it can perpetuate notions of Whiteness as the standard, positioning the experiences of anyone who is not White as "other". Grouping racial identities also perpetuates perceptions that experiences of all groups within the "BIPOC" category in STEM are homogenous, when this is not likely the case. For example, STEM students report different frequency and type of microaggressions experienced based on racial group (Lee et al. 2020). However, comparisons between White and BIPOC respondents can still be informative because BIPOC communities often share the experiences of discrimination and underrepresentation more than they do with the majority population (e.g., Malcom and Malcom 2011; Ong et al. 2011; American Institute of Physics 2020; Barthelemy et al. 2022).

Although categorizing respondents broadly into BIPOC and White groups does not acknowledge the nuanced and varied experiences within BIPOC communities, we make this simplification at times for three central reasons: (1) to somewhat equalize group sizes and descriptively compare White and BIPOC respondents in a predominantly White sample, (2) to advance research that has evidenced the prevalence of race-based discrimination in physics (e.g., Malcom and Malcom 2011; Ong, et al. 2011; American Institute of Physics 2020), and (3) to provide foundational results to the physics community that can serve as rationale for pursuing identity-specific research (i.e., focusing on Black physicists).

To obtain insights about the demographics of physics students and physicists in Canada, respondents were asked to specify gender identity, sexual identity, age bracket, race/ethnicity, disability status, first language, and immigration status. Respondents were asked to identify which province or territory they primarily lived in, primary position(s) in academia or the private sector, highest education level, education field, future plans, and research field. The results section that follows only includes analyses pertaining to respondent demographics, with a focus on students as they comprised the largest proportion of respondents. Additional demographic details appear in Supplementary Materials.

The survey used a self-report methodology, consistent with previous research including the Canadian Census (Statistics Canada 2020). We adapted wording from the Canadian census for several questions (e.g., race or population group), given phrasing has been vetted by Statistics Canada demographers and shown to produce quality data (Statistics Canada 2020). By offering the option to self-describe for all identity categories, we were able to gather any information beyond what is possible with only closed-ended response options and obtain suggestions for improvements of this question in future research.

# Results and discussion

## Analysis

Responses were retained for analysis if respondents completed the study in full. Duplicate cases were removed ($n = 9$) prior to analysis, leaving a final sample of $n = 2532$. In the following, we descriptively compare the demographics of students and professionals to build a picture of the future of the Canadian physics community and help future researchers and decision-makers identify strategies to promote diversity and inclusion. We focus on the student respondents as they constituted the largest proportion of survey respondents (i.e., 895 undergraduate students, 634 graduate students). In some instances, we make descriptive comparisons of the survey data to the Canadian population using information from the Canadian Census. However, as in other similar work (Aldosary et al. 2023), no statistical tests comparing the sample data to the census data were conducted. Instead, usage of "over" and "under" representation refer to descriptive differences between the sample data and Canadian population.

**Fig. 1.** This figure displays proportions of Indigenous, Black, POC, and White respondents, shown in total as well as split between students and professionals. Note: POC = Chinese, Japanese, Korean, West Asian, Southeast Asian, Arab, Latin American, Filipino, South Asian, Multiple selected (multiracial), and those who preferred to self-describe and indicated a BIPOC identity. Black = those who selected only Black. Indigenous = those who selected only Indigenous.

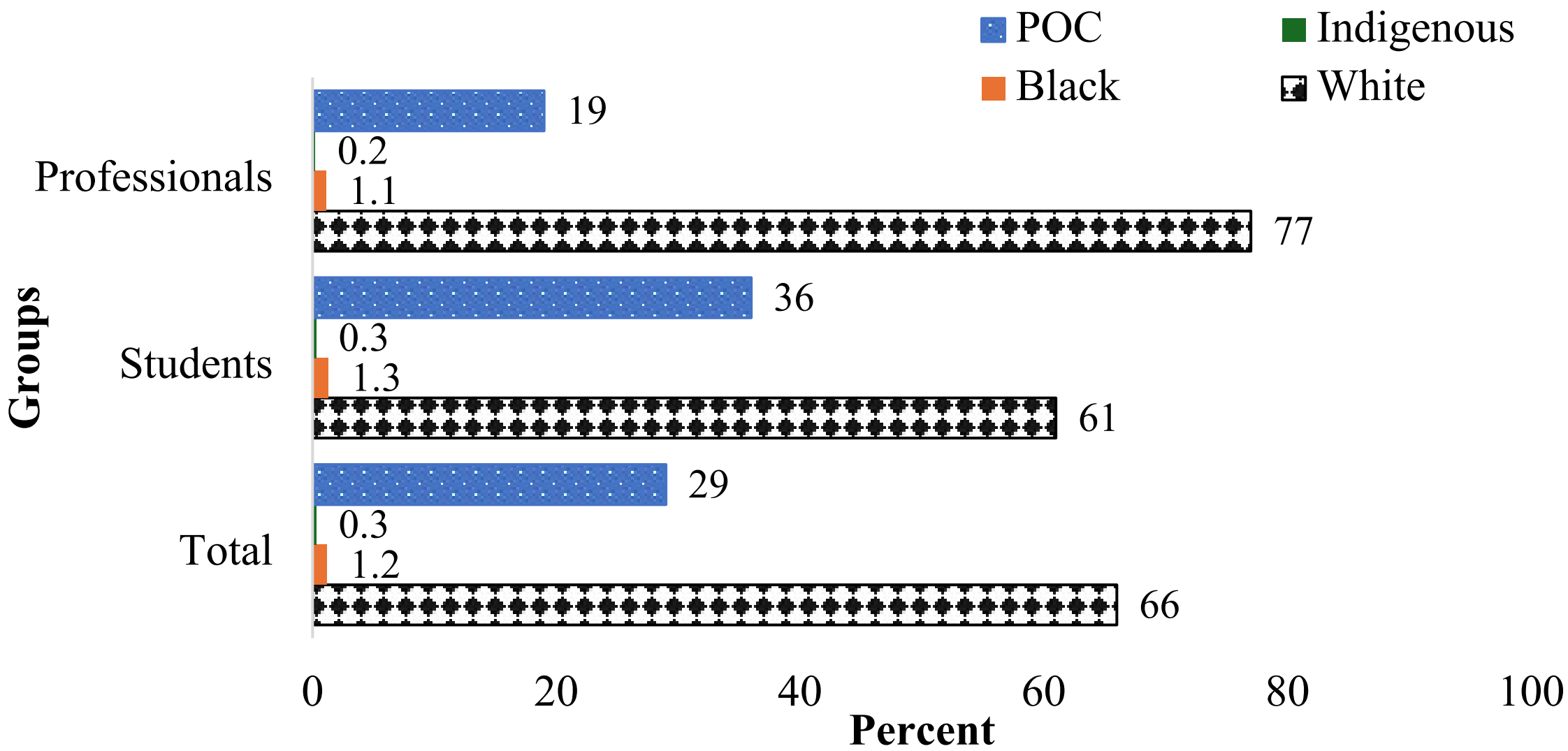

## Location and field

Survey responses were received from every region of Canada, including a small number from Northern Canada. Self-reported locations were Ontario (42.2%), Quebec (17.7%), British Columbia (16.2%), Alberta (9.2%), Nova Scotia (5.6%), Saskatchewan (2.3%), Outside Canada (2.3%), Manitoba (1.5%), New Brunswick (1.5%), Newfoundland (1.1%), Prince Edward Island (0.3%), Northwest Territories (0.1%), and Did Not Answer (0.1%). Responses were received from over 75 physics departments, research institutes, and from those in the private sector and government. When asked about field of education, most respondents selected multiple areas (37.6%), followed by physics (general; 22.7%), reflecting the multidisciplinary nature respondents' education paths. A selection of the main demographics is presented here, with additional data included in Supplementary Materials.

## Black and Indigenous physicists

This survey provided the first data in Canada pertaining to the participation of Black and Indigenous students and researchers in physics. Just 1.2% of respondents identified as only Black, while 0.3% identified as only Indigenous. Another 1.1% identified as Indigenous and another racial identity, and another 0.4% identified as Black and another racial identity. According to Statistics Canada (2023), 4.3% of the Canadian population is Black, and 5% is Indigenous (Government of Alberta 2023). Based on current data, we find a severe lack of representation of Black and Indigenous physics students and physicists among respondents in Canada. Figure 1 shows the racial identities of respondents in the total sample, and among students and professionals; respondents in both groups overwhelmingly identified as White. These data are consistent with a survey of the Canadian Organization of Medical Physicists that reported Indigenous representation of less than 1%, and Black representation of 2% (Aldosary et al. 2023), and with other Canadian data showing low representation of Black and Indigenous students and faculty members, support staff, and admin staff in several Canadian Universities (CAP 2023).

We draw out specific statistics for Black and Indigenous respondents because these groups are the least represented overall in physics and may have unique experiences and barriers to pursuing physics in the current landscape (AIP 2020; Herkimer 2021). Post-secondary education is a part of an interconnected system that reproduces and amplifies social, historical, and racial inequities that continue to impact the schooling and careers of Black and Indigenous Canadians. For example, Black students are disproportionately streamed into Applied and Essentials programs of study as early as Grade 10, which significantly impacts their ability to pursue post-secondary education, particularly within STEM fields (James and Turner 2017). Further, Black and Indigenous physics students rarely see themselves represented among teachers and faculty, which significantly decreases the likelihood of passing courses and pursuing a university degree (Gershenson et al. 2017; CAUT 2018; Odle et al. 2022). Indigenous students are more likely to leave post-secondary education early, with one U.S. study finding that only 8% of these cases are due to academic failure, indicating how education systems fail to adequately support Indigenous students in these environments (Martinez 2014; Bernardo et al. 2016). Future research should explore the unique experiences of Black and Indigenous physics students so that actions to improve representation are data informed.

**Fig. 2.** This figure displays proportions of men, women, and gender diverse respondents, shown in total and split between students and professionals. Gender diverse = those who selected two-spirit, nonbinary, transgender, genderqueer, gender nonconforming, and multiple gender identities.

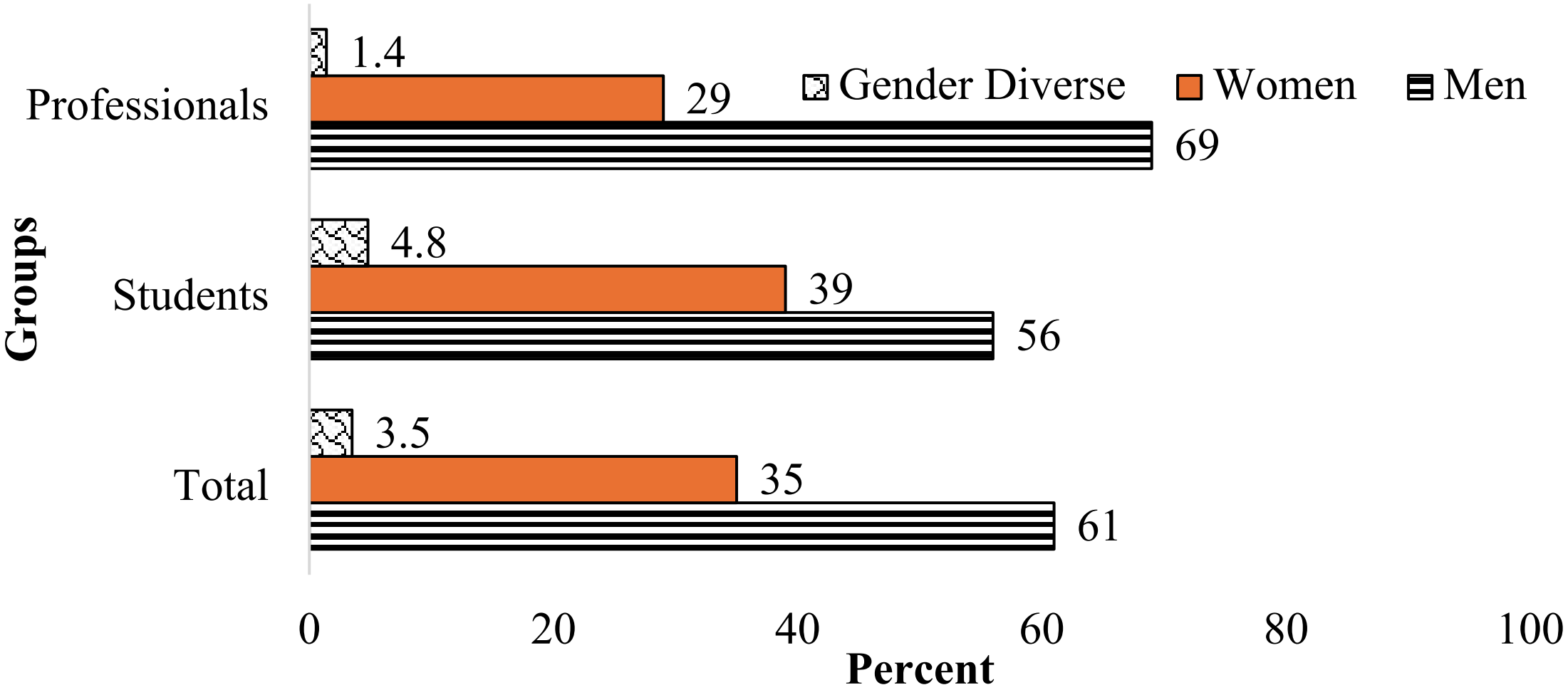

## Gender diversity in physics

Our survey also provided the first Canadian data on the representation of gender-diverse students and professionals in the physics community (i.e., nonbinary, two-spirit, genderqueer or gender nonconforming, transgender, and open-ended self-described gender identities). Some past research on gender-diverse people in physics has been conducted in the United States and focused largely on experiences (e.g., Barthelemy et al. 2022). Data collected in Canada on gender diverse identities in physics are limited as most institutions only collect binary gender data (men and women; CAP 2023).

Figure 2 displays gender identity for respondents in the total sample, and split by students, and professionals. Gender diverse people comprised a larger proportion of our dataset (3.5% of the total sample) than in the general population in Canada (0.33% are transgender or nonbinary; Statistics Canada 2022*a*; 2021 census), particularly among students; however, different ways of measuring gender identity make direct comparisons difficult. Although youth in Canada are more likely to be gender diverse than other age groups (0.79% of Canadians aged 15–24 years identified as nonbinary or transgender; Statistics Canada 2022*a*), we find an abundance of gender diverse students in the current sample. This may be in part due to sampling procedures (i.e., snowball recruitment, self-selection). Another factor may be that the current survey had six gender options and an option to self-describe, and the 2021 Canadian Census had a question about sex assigned at birth, and then a question on gender with two options (male and female) and the option to self-describe (Statistics Canada 2022*b*). Recent research (Casper et al. 2022) showed that surveys with more gender options reveal more gender diversity, perhaps due to an increased sense of safety signalled by language and/or the subsequent ease of responding.

While women comprised a moderate proportion of respondents, men constituted the largest proportions of student and professional respondents, despite being just under half of the entire Canadian general population. According to data from Statistics Canada, in 2021, cisgender men comprised about 49% of the general Canadian population, and cisgender women comprised about 51% (Statistics Canada 2021*a*). Our data are consistent with international statistics showing the overabundance of men in physics (American Institute of Physics 2022; Institute of Physics 2022), and other Canadian samples (Aldosary et al. 2023; CAP 2023). Among students, close to 5% identified as gender diverse, and 1.4% among working professionals.

## Intersectional analyses

We assessed the proportions of White and BIPOC respondents along dimensions of gender identity. As illustrated in Fig. 3, White men were the majority of respondents among both students and professionals. However, among student respondents, there was greater representation of diverse and intersecting identities compared to professional positions. For example, 34% of students identified as White men. Among physics faculty (not all working professionals), 64% of survey respondents identified as White men. According to Statistics Canada, close to 70% of the Canadian population is White (Statistics Canada 2022*c*). In this sample, the proportion of White women was roughly consistent across students and working professionals (24% vs. 22%, respectively), whereas the proportions of BIPOC men, BIPOC women, and gender-diverse physicists was notably lower among working professionals than students. For example, among students, 22% identified as BIPOC men compared to 13% among working professionals, and 15% identified as BIPOC women compared to 7% of working professionals, representing close to a 2:1 ratio.

**Fig. 3.** This figure displays proportions of gender identities and racial groups, shown in total as well as split between students and professionals. BIPOC = Black, Indigenous, and People of Colour. Gender diverse = those who selected two-spirit, nonbinary, transgender, genderqueer, gender nonconforming, and multiple gender identities.

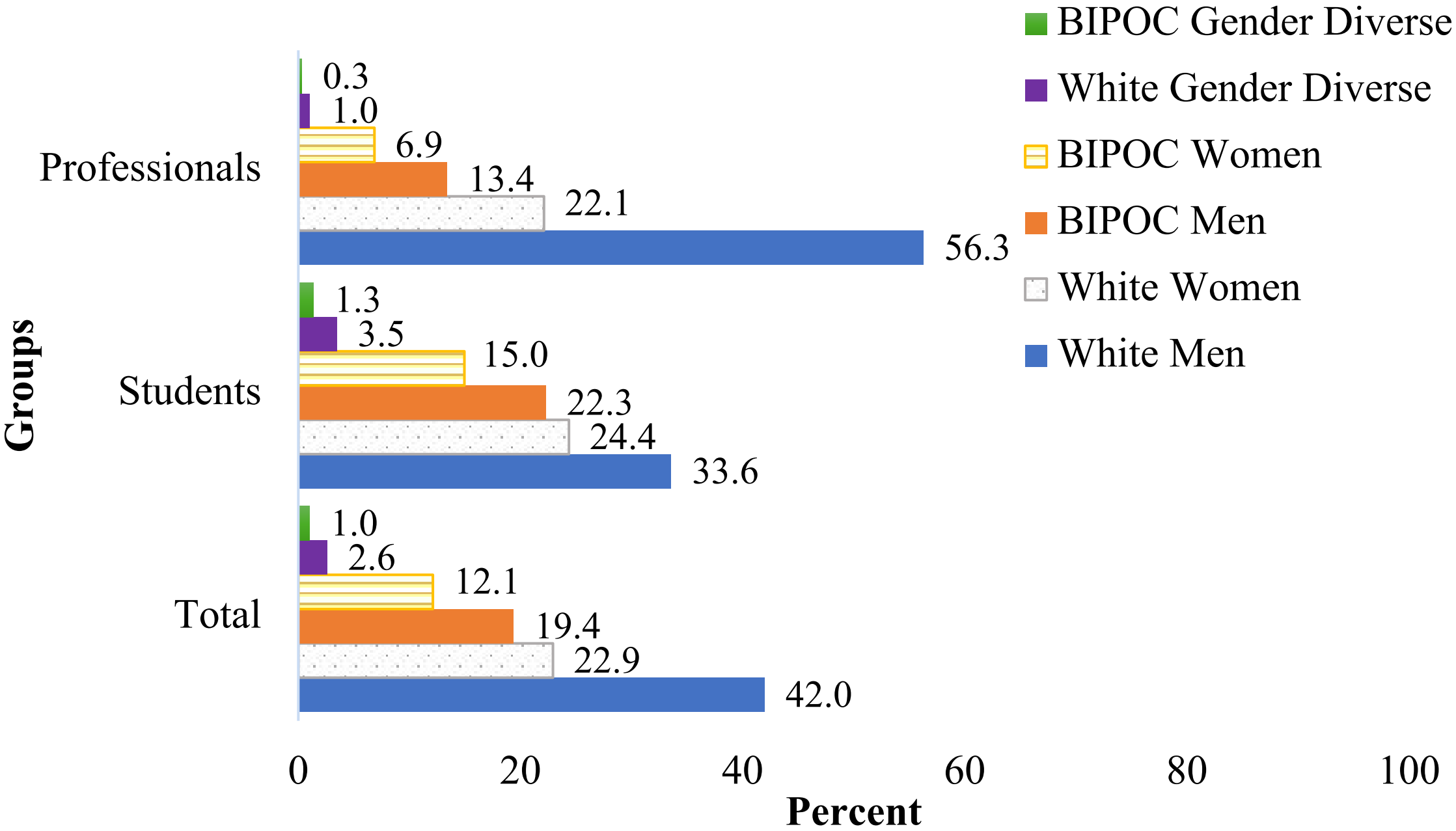

As illustrated in Fig. 4, BIPOC representation fell precipitously (by one-half) between student to professional roles, while White women's representation remained relatively consistent across roles. The data highlights the need to identify and remove the associated barriers to ensure the diversity of those pursuing physics degrees matches the future professionals employed in the field.

Given the current demographics of students who could reach the senior ranks of physics faculty members in two decades, while there will be more demographic diversity overall, there will likely not be proportional representation in physics even 20 years from now. Indeed, projections from Statistics Canada estimate that by 2041, the BIPOC population could account for 38%–43% of the Canadian population (Statistics Canada 2022*d*). In 2021, this proportion was approximately 30% (Statistics Canada 2022*c*).

## Sexual orientation

Another important first for this research was the collection of data on sexual orientation among physics students and physicists in Canada. As shown in Fig. 5, most respondents identified as heterosexual (80%), both among students (74%) and professionals (89%). However, we found notable differences in the proportions of those who identified as sexually diverse (i.e., asexual, bisexual, gay, lesbian, pansexual, questioning, queer, and prefer to self-describe) between students and professionals. Among students, there were more than twice as many sexually diverse respondents (26% vs. 11%, respectively).

Our findings are somewhat expected, given that nationally, almost one-third of self-identified 2SLGBTQIA + Canadians are under 25 years old (Statistics Canada 2021*b*). Among professionals, the current survey also shows a larger representation of sexually diverse people compared to the general Canadian population, in which 4% identify as part of the 2SLGBTQIA + community (Statistics Canada 2022*e*). At the intersections of sexual orientation, gender identity, and race, our data show that women in physics comprise the majority of sexually diverse respondents, and that this is consistent across racial groups. Among both White women and BIPOC women, sexually diverse respondents were more than three times the proportion found in the general Canadian population (Statistics Canada 2022*d*).

When sexually diverse respondents are further disaggregated (Fig. 6), we find that the majority identified as bisexual, and this finding held for BIPOC and White women (12% in both groups). We also find that among students, bisexual women (15%) far outnumbered bisexual men (4%); a finding in line with, and even more pronounced, than recent Statistics Canada data showing that bisexual women (332, 000) outnumbered bisexual men (161, 200) approximately 2 to 1 (Statistics Canada 2021*b*) in the general population. The strong representation of bisexuality among physicists has also been shown in data collected via the Institute of Physics

**Fig. 4.** This figure displays proportions of gender identities and racial groups, shown across roles from along the academic track and in nonacademic roles. BIPOC = Black, Indigenous, and People of Colour. Gender diverse = those who selected two-spirit, nonbinary, transgender, genderqueer, gender nonconforming, and multiple gender identities. See Supplementary Materials tables for exact group values.

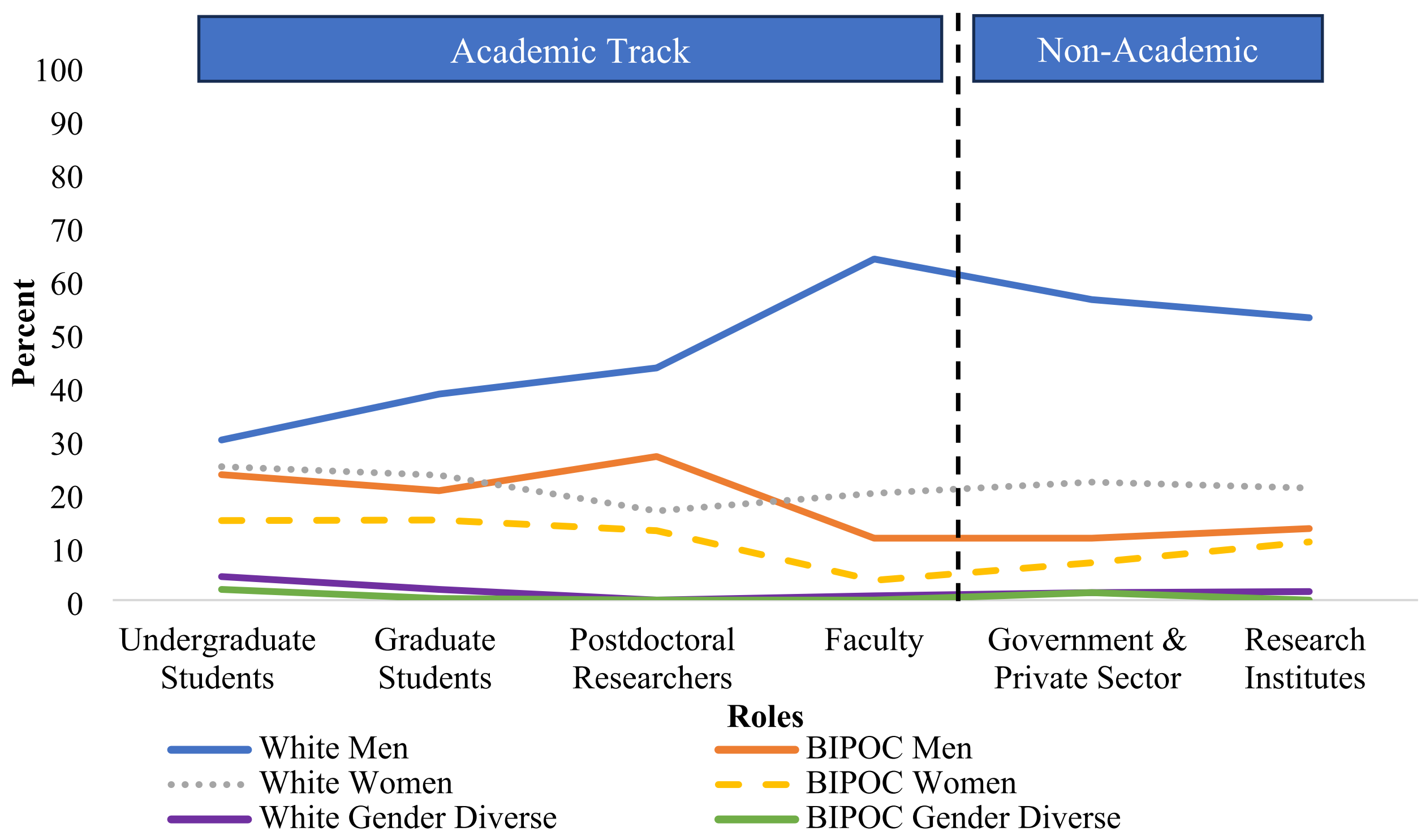

**Fig. 5.** This figure displays proportions of sexually diverse and heterosexual respondents, shown in total as well as split between students and professionals. Sexually diverse = those who selected asexual, bisexual, gay, lesbian, pansexual, queer, questioning, prefer to self-describe, and multiple sexual orientations.

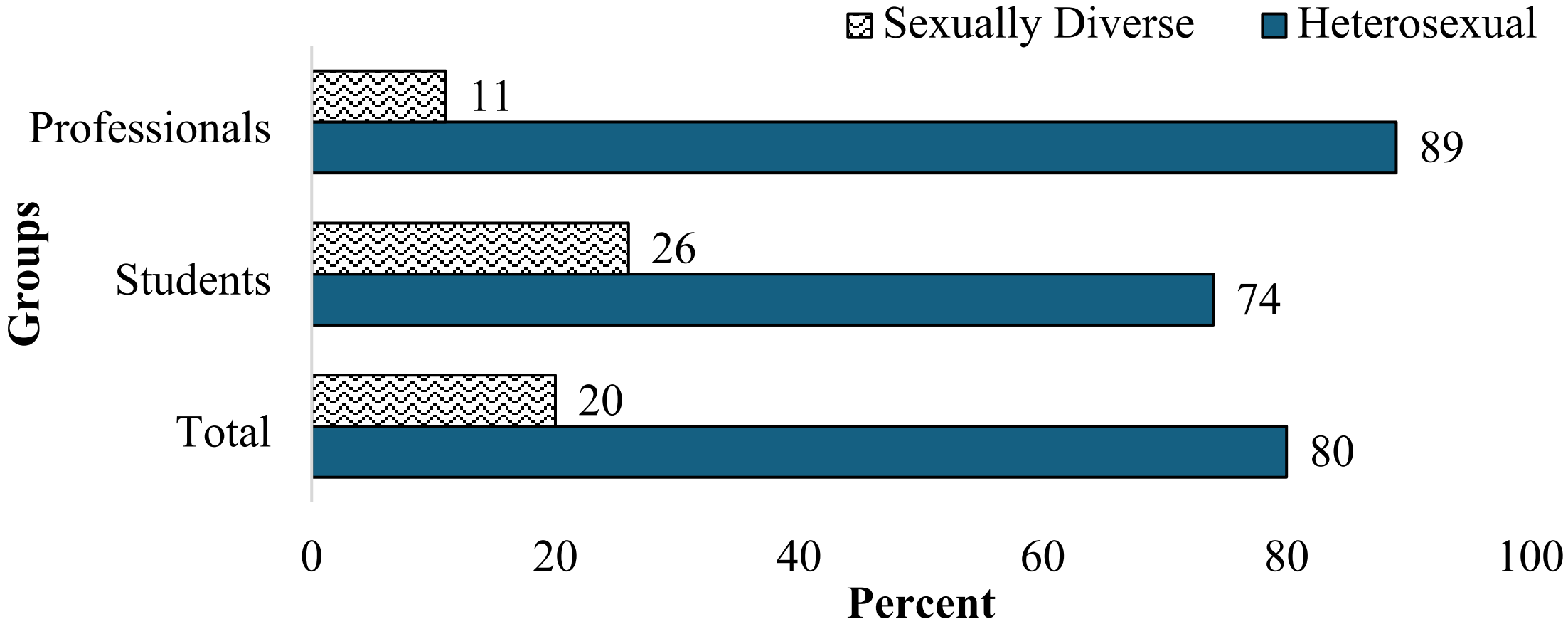

Member Diversity Survey, such that in 2019, 11% of respondents identified as 2SLGBTQIA+, with bisexuality having the largest representation (6%; Institute of Physics 2022).

## Disability

We also collected some of the first data on disability within the Canadian physics community. When asked about disabil-ity, 7% of respondents identified as disabled[1]. Other data collected in Canada also showed a low rate of reported disability among physics students (6.9% of Bachelor students) and Faculty members (0.5%–2% depending on rank; CAP 2023). However, Aldosary and colleagues (2023) found higher rates (23%) of disability in their COMP membership, which could be because their survey was mandated for members, or an artifact of differences in measurement or definition of disability. Among students, the proportion of disabled respondents was 8%, and among professionals it was 5%. According to the Canadian Survey on Disability (Statistics Canada 2023), approximately 22% of working-age Canadians (25–64) reported having a disability, suggesting that the current data reflects a low level of representation of disabled people in the Canadian physics community. Figure 7 shows the proportion of respondents who reported being disabled across gender and racial identities. Overall, we find that the proportions of disabled people are largest among those with diverse gender identities, particularly among White gender diverse students and professionals. Notably, a quarter of respondents from BIPOC gender diverse backgrounds identified as being disabled, underscoring the importance of disaggregating data along multiple dimensions of identity. Women were more likely than men to identify as disabled among both students and working professionals, a finding consistent with the Canadian Survey of Disability (Statistics Canada 2023). Across all but one group of respondents (i.e., White gender diverse), students were more likely to identify as disabled than professionals, which could reflect a greater willingness to identify as disabled among younger cohorts, but could also be a result of those with a disability leaving physics, or having fewer opportunities to pursue physics in the past due to inaccessibility; additional data are required to tease apart these hypotheses.

**Fig. 6.** This figure displays proportions of respondents' gender identities, racial groups, and sexual orientation split between students and professionals. Sexually diverse = those who selected asexual, bisexual, gay, lesbian, pansexual, queer, questioning, prefer to self-describe, and multiple sexual orientations. BIPOC = Black, Indigenous, and People of Colour. Gender diverse = those who selected two-spirit, nonbinary, transgender, genderqueer, gender nonconforming, and multiple gender identities.

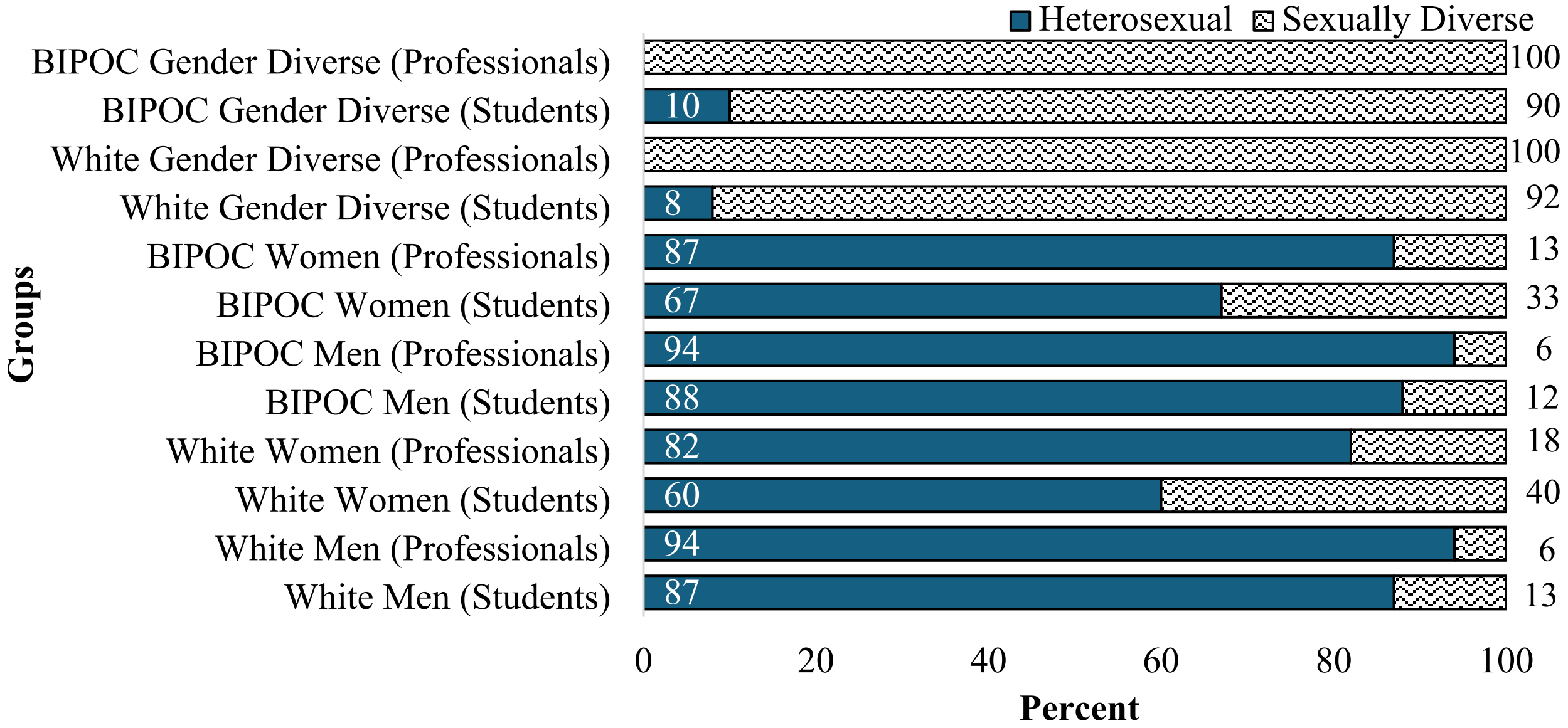

Figure 8 shows the proportion of respondents with a disability across sexual orientations among students and professionals. Here we find that the proportion of disabled respondents is largest among sexually diverse students, and this proportion is somewhat consistent with sexually diverse professionals. Similarly, the proportion of disabled respondents is comparable among heterosexual respondents, both students and professionals. The current data therefore point towards an important intersection of sexual diversity and disability that should be further researched in the Canadian physics community. Indeed, we find that the proportion of sexually diverse students who reported having a disability was more than three times higher than the proportion of heterosexual students with a disability.

Among those who identified as disabled, 5% reported receiving full accommodations at their place of work or study, whereas 25% reported receiving adequate accommodations, 14% reported having received partial, inadequate accommodations, 3% chose to enter a free-form text response with additional detail, and 53% reported they had not sought any accommodations. Among students, the proportion of those who reported receiving full accommodations was larger than among professionals (5.5% vs. 4.5%, respectively).

[1] We recognize the ongoing discourse about using identity-first versus person-first language in relation to disability. Guided by findings from a recent study showing that identity-first language was preferred by 49% of disabled people across 23 countries (https://dl.acm.org/doi/pdf/10.1145/3517428.3544813) we use the words "disabled people" in this report.

**Fig. 7.** This figure displays proportions of respondents' gender identities, racial groups, and disability split between students and professionals. BIPOC = Black, Indigenous, and People of Colour. Gender diverse = those who selected two-spirit, nonbinary, transgender, genderqueer, gender nonconforming, and multiple gender identities.

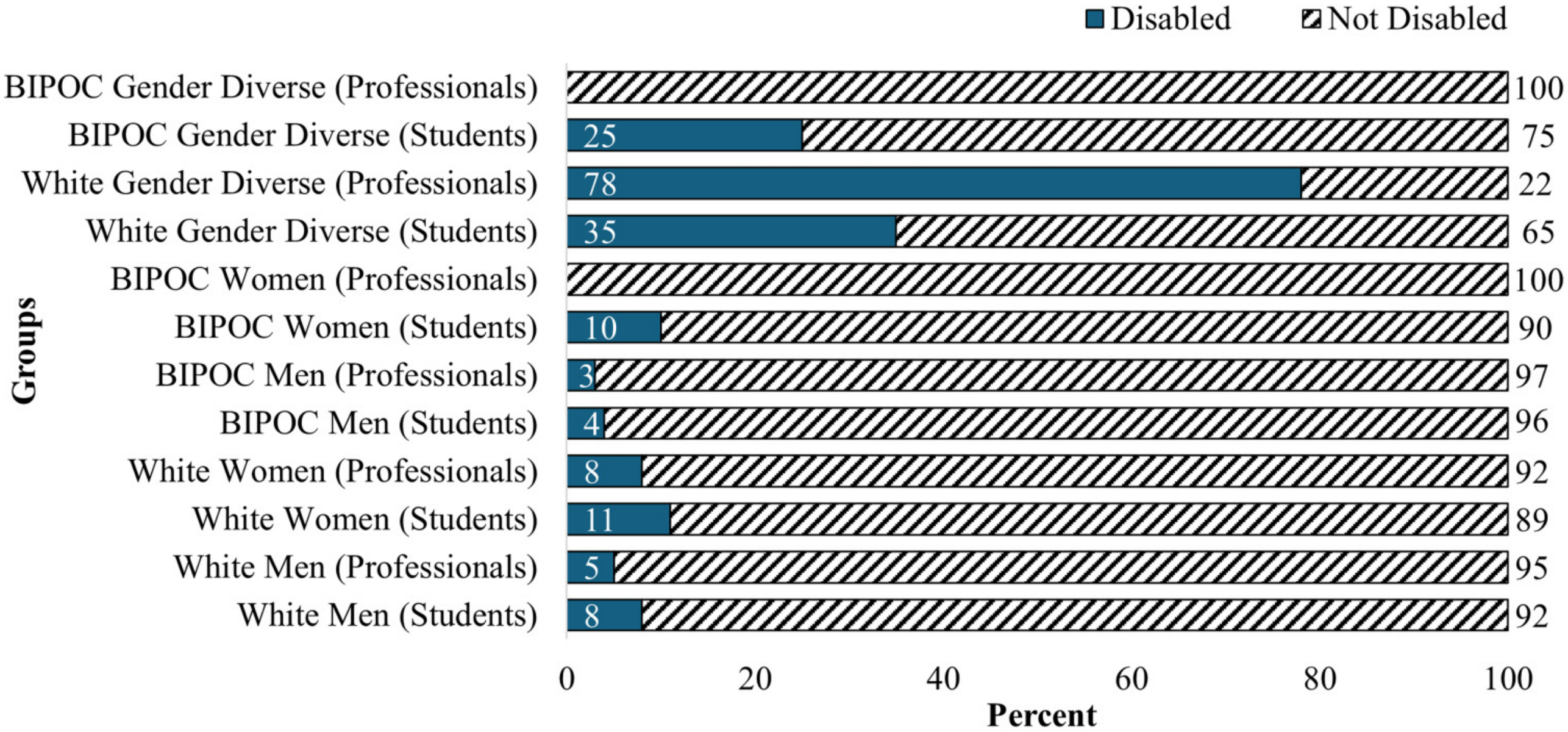

**Fig. 8.** This figure displays proportions of respondents' sexual orientation and disability split between students and professionals. Sexually diverse = those who selected asexual, bisexual, gay, lesbian, pansexual, queer, questioning, prefer to self-describe, and multiple sexual orientations. Disabled = self-identified as disabled.

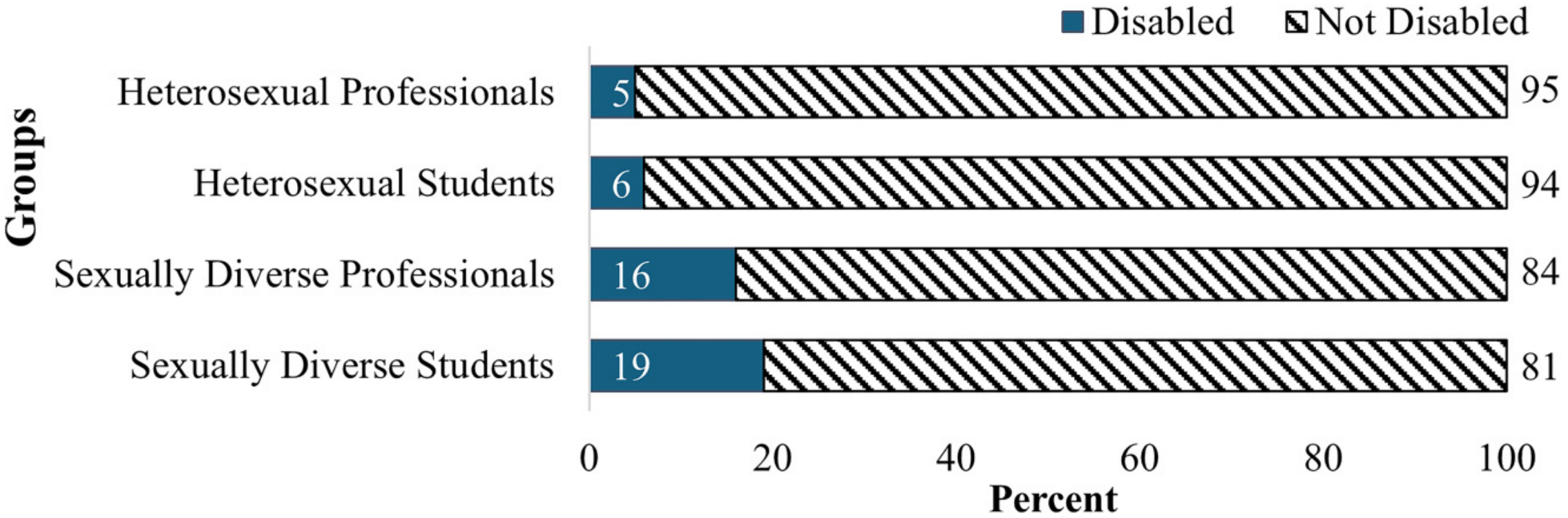

In 2022, 35% of employed Canadians with disabilities aged 25–64 years needed at least one workplace accommodation (Schimmele et al. 2024) to be able to work. Our results show low rates of accommodation requests, which could reflect underrepresentation of more severe disabilities (which are more likely to require workplace accommodations), a lack of systems to facilitate such requests, a lack of safety to disclose disability, a belief that requests may not be granted, or some combination of these factors. It must be considered, however, that some disabled people do not request workplace accommodations despite potentially benefiting from them; often due to discomfort asking (42%) or fearing a negative outcome (34%; Statistics Canada 2017). While buildings and workplaces are becoming more accessible in general, disabled students and working professionals likely continue to encounter challenges requesting accommodations at their place of work or study—and physics environments are no exception. The need to make physics more accessible has resulted in calls to action for proactive measures among physics educators (Dounas-Frazer et al. 2022), physics laboratories (Caden et al. 2024), and recommendations for increasing accessible science labs (Sukhai et al. 2014).

Given the national scope of the survey, it is important to also highlight responses from Canada's Francophone community. Of all respondents, 10% completed the survey in French—considering that 22% of the Canadian population speaks French as a first language, French speakers were underrepresented in our study (Canadian Heritage 2024). Most Francophone respondents (84%) were located in Quebec. We found similar distributions of gender identity, disability, and sexual orientation between those who completed the survey in English versus those who completed the survey in French (See Table 4 in Supplementary Materials). Notably, the proportion of BIPOC respondents was close to three times larger among English speakers (33.3%) than among French speakers (10.3%).

## Conclusions, limitations, and future research

The present research provides a wealth of information about the physics community in Canada that was previously unknown. With responses from every region in the country and a robust sample of respondents, this work revealed several key findings. It is notable that the majority of respondents were students, which suggests that students in physics demonstrated a higher degree of buy-in when it came to thinking about their identities, as well as being able to devote the time and effort required to participate. Indeed, the physics students of today are the physicists of tomorrow, so the fact that student respondents were motivated to participate in a voluntary equity, diversity, and inclusion survey indicates an openness to inclusion work that we hope persists into their professional and academic trajectories. Given that physics students value empathy and identity (Stiles-Clarke and MacLeod 2018; American Institute of Physics 2020), that Canadian post-secondary institutions are working to improve inclusion (Universities Canada 2023), and funding agencies are embedding equity, diversity, and inclusion into their programs (Government of Canada 2023), it is incumbent upon researchers to embrace inclusive principles in their scientific practice.

In addition to a high level of responsiveness among students, our research also showed greater demographic diversity among students versus professionals. This difference could be due to a multitude of factors. For example, it could be that the efforts to make the physics environment more inclusive have been successful, or this could reflect the growing demographic diversity of students in post-secondary education in Canada in general, such as through the increase in international students (Choi et al. 2021).

Our research highlights a serious lack of Black and Indigenous respondents, underscoring the importance of specific recommendations put forth in frameworks like the Scarborough Charter to support the recruitment, retention, and flourishing of Black scholars (University of Toronto 2022). In the present work, when we disaggregated the BIPOC data to focus on Black and Indigenous respondents specifically (i.e., BI), we found stark differences in their representation compared to POC, challenging the notions of BIPOC homogeneity. Indeed, there is evidence that not all BIPOC people are represented or rewarded equally in science; for example, research shows that Asian men are overrepresented in science jobs and tend to earn higher salaries than other racial groups (Fry et al. 2021). We therefore recommend that future researchers adopt a similar disaggregating approach when analysing data about the physics community in Canada, so that the most notable disparities are evident.

The present research also revealed gender diversity among respondents in physics, particularly among students. Despite this diversity, White men still comprised the largest proportion of respondents, which is not unique to the Canadian physics context (see Bruun et al. 2018; American Physical Society and the Integrated Postsecondary Education Data System 2022). White men play an important role in striving to make science more inclusive as they comprise the majority of the physics community (including most senior leadership roles) and are therefore well-positioned to advocate for institutional change. White men who are more senior and established in their physics careers also have an opportunity to leverage power and may face fewer consequences for doing so; men who confront sexism are evaluated more positively (vs. women), and their confrontations are seen as more legitimate (Drury and Kaiser 2014). We invite White men in physics (and all who have power and privilege) to be active in inclusion initiatives, so that efforts to build a more accessible physics community in Canada include all groups, including allies and/or accomplices. If all physicists focus efforts on how to contribute to meaningful, systemic change, it is possible for the physics community to become a more just and equitable environment.

An interesting finding in the current research was the relatively consistent representation of White women among students and professionals. It could be that such representation is a result of successful employment equity interventions, in line with the assertion that White women have benefited more than other demographic groups from employment equity and/or affirmative action programs and policies in North America (Wise 1998). This finding highlights the importance of rethinking current initiatives and developing more intersectional strategies that will also benefit BIPOC women or women with different intersecting identities in science. If earlier interventions were partially successful in increasing the accessibility of physics to White women, it is possible and necessary to improve on this earlier work and further open the field to BIPOC, disabled, and/or gender diverse physicists, and most importantly—retain them. We call for solidarity from White women, who may or may not hold positions of power in physics, as they possess privilege in their Whiteness, but also potential disadvantage because of their gender.

Another noteworthy contribution of the current study is data about sexual orientation in the physics community in Canada. Interestingly, we found that among respondents, sexually diverse people were far more represented compared to the national average, especially among students. When further parsing apart this result, the data revealed that the largest proportion of sexually diverse respondents were bisexual women. In general, research shows that people have more positive attitudes towards bisexual women than bisexual men (Friedman et al. 2014), so this result could be attributed to women feeling safer to disclose their sexual identities than men; however, this is merely speculative. Further research is needed to truly understand if and why the physics community has a larger proportion of sexually diverse people than the national average. Our research also draws attention to the need for intersectional analyses of physics identities. For example, those at the intersections of disability, gender diversity, and sexual orientation report unique and overlapping experiences of oppression (Alimi and Abbas 2019). If we are to strive for the inclusion of diverse voices in physics in Canada, our efforts must not only include but prioritize the safety and participation of the most marginalized.

## Limitations

Although our results provide valuable new insights, the present study has limitations. For example, the data included in this report were collected in part by using a snowballing recruitment method, which limits the generalizability of findings as responses reflect the identities of a subset of volunteers who chose to respond. Thus, it is imperative that more systemic data be collected in Canada to ensure that information accurately reflects the demographics of the physics community. That being said, comparisons to other data including a survey of the Canadian Organization of Medical Physicists members (Aldosary et al. 2023), the Canadian Association of Physicists (CAP 2023), and the Institute of Physics (2022) found similar proportions of women (Aldosary et al. 2023; CAP), diverse sexual identities (Institute of Physics 2022), and Black and Indigenous physicists (Aldosary et al. 2023; CAP 2023) as we observed in the current research. However, the current data showed higher proportions of gender minorities than would be expected based on the limited past surveys of physicists that included gender identities beyond men and women (CAP 2023) and the Canadian census (Statistics Canada 2021*a*).

In the present study we received very few responses from those employed in industry or government. It is possible that this is because we utilized the CAP to recruit members and this organization tends to be comprised mostly of academic institutions. Our projected sample size for government and industry was based on older information (while outdated, these sources are the most recent from the Canadian context; Robertson and Steinitz 1997; Strickland 2017). A report from the Government of Canada (Bonikowska et al. 2022) found that the highest number of doctoral graduates who worked outside academia obtained their degree in the broad area of physical sciences and technology (30.3%); however, this statistic includes chemistry, biochemistry, biophysics, molecular biology, physics, biology, and microbiology making it difficult to parse out only those physicists working outside of academia in the Canadian context. Future research would benefit from obtaining broader representation across sectors so that results more comprehensively reflect the physics community as opposed to primarily representing the academic sector.

Finally, some studies have used inferential tests to statistically compare sample data from physics communities and census data to test whether the proportions of different identities in physics (e.g., women) are statistically different from the population proportions (Casper et al. 2022; Rankin et al. 2022). However, other studies (Aldosary et al. 2023) have compared sample data with census data descriptively. We chose a similar descriptive approach. The purpose of comparing our sample data with the Canadian population was to provide an illustration of who is participating in physics in Canada and who is not. We hope to encourage researchers, policymakers, and others in seats of power to go beyond the numbers, using our findings in combination with experimental and experiential data to inform policy and institutional actions.

Despite these limitations, we are confident that the present findings will be useful and informative for educators, researchers, and policy makers, as well as physics students and physicists in Canada. Our study went far beyond previous surveys that have focused on a particular segment of the Canadian physics community, and gathered responses from physicists across Canada, across fields, industries, and points in their career. We collected comprehensive data on race, disability, gender, and sexual orientation by asking physicists about who they are, digging deeper into identity than previous research relying on institutionally collected data with limited response options. Assessing representation as we have done here is an important starting point, from which we hope to promote well-resourced, up-to-date demographic data collection on a national level. To advance authentic inclusion work, we hope to inspire further research into demographic representation and experiences in the Canadian physics community in the interest of improving access and participation. We welcome inquiries and collaborations from researchers or organizations interested in building on this work. A recent examination (Stiles-Clarke and MacLeod 2025) of belonging in a research-intensive institution in Canada suggests that departments create mentoring programs as a means of enhancing belonging, and ultimately retention. We advocate for expanding public and institutional investment in equity, diversity, and inclusion research in science—both in Canada and internationally—as a critical foundation for meaningful, evidence-based change in our communities, as we strive to create more inclusive environments for future generations of physicists in Canada. As North America's shifting socio-political climate poses challenges to inclusion efforts (e.g., defunding EDI-related grants in science), understanding and supporting diverse scientists—especially physicists—remains vital to driving and sustaining scientific innovation.

## Acknowledgements

Hewitt acknowledges that his work is done in Kjipuktuk (Halifax), which is located in Mi'kma'ki, the ancestral and unceded territory of the Mi'kmaq People. This territory is covered by the "Treaties of Peace and Friendship" which Mi'kmaq Wəlastəkwiyik (Maliseet), and Passamaquoddy Peoples first signed with the British Crown in 1726. The treaties did not deal with the surrender of lands and resources but in fact, recognized Mi'kmaq and Wəlastəkwiyik (Maliseet) title and established the rules for what was to be an ongoing relationship between nations on Mi'kma'ki.

Smolina acknowledges that her work is done at The University of Toronto, the Hospital for Sick Children, and the Mouse Imaging Centre, which reside on the traditional lands of the Huron-Wendat, the Seneca, and the Mississaugas of the New Credit First Nation. These territories are protected by the "Dish with One Spoon" wampum agreement, which all guests of this land have a responsibility to honour and uphold.

The WinS team acknowledges that their work is done on the shared traditional territory of the Neutral, Anishnaabe,

and Haudenosaunee peoples. This land is part of the Dish with One Spoon Treaty between the Haudenosaunee and Anishnaabe peoples and symbolizes the agreement to share, protect our resources, and not to engage in conflict. This project was support by the Natural Sciences and Engineering Research Council of Canada.

## Article information

### Editor(s)

Paul Dufour, Yann Joly

### History dates

### Notes

The article was originally published with errors that have now been corrected. See Correction: https://dx.doi.org/10.1139/facets-2025-0413.

### Copyright

### Data availability

The dataset generated by the survey research during and/or analyzed during the current study are not available because doing so would violate our Dalhousie Research Ethics Board approval. This approval states that only Hewitt, Smolina, Tassone, and Hennessey will have access to the anonymized data.

## Author information

### Author ORCIDs

Eden J. Hennessey https://orcid.org/0000-0001-6529-313X
Anastasia Smolina https://orcid.org/0000-0002-7162-422X
Adrianna Tassone https://orcid.org/0000-0002-5076-9104
Shohini Ghose https://orcid.org/0000-0001-6214-4518
Kevin Hewitt https://orcid.org/0000-0002-6336-0789

### Author notes

Eden J. Hennessey and Anastasia Smolina are joint first authors.

### Author contributions

Conceptualization: AS, KH
Data curation: EJH, AS
Formal analysis: EJH, AT, AJ, SG
Investigation: AS, KH
Methodology: EJH, AS, KH
Project administration: AS, SH, KH
Resources: EJH, SH
Software: EJH
Supervision: SG, KH
Validation: EJH
Visualization: EJH
Writing – original draft: EJH, SG
Writing – review & editing: EJH, AS, SH, AT, KH

### Competing interests

The authors declare there are no competing interests.

## Supplementary material

Supplementary data are available with the article at https://doi.org/10.1139/facets-2025-0126.